# An innovating approach to teaching applied to database design. Improvement of Action Learning in Lifelong Learning.

C Béchade
*University Institute of Technology – Angers-Cholet*
*University of Angers, France*

**Abstract**

For now 10 years, the Action Learning has allowed employees of University of Angers, private and public Companies to be initiated with the design of database, on projects financed by professional structures. These innovating training periods are carried out within the framework of the University College of Further Education of the University of Angers.
Database design is a process initially reserved to the professional data processing specialists, coming from French Level-2 technological courses (2-year degrees) or Engineer Schools (Master).
The pedagogical model of technological courses has integrated for more than 20 years transverse semester projects, in order to give the students the opportunity to apply newly acquired knowledge, coordinated by teachers. Action Learning requires teachers to assume the role of supervisors for the project management.
The objective of Action Learning is to transmit not only knowledge from teachers, but also the experience of consultants to trainees having no competence in data processing, but who have the knowledge of their business process.
The present paper shows that Action Learning puts together the factors for success of French technological courses, the adaptability of pedagogy provided to the vocational training, and finally the competence of service provider,
Keeping the best parts of those three complementary approaches makes it possible for this kind of formation to achieve teaching and professional, assessable and long lasting goals.
Action Learning belongs to the French policy that aims to improve the volume and the quality of the contracts between Universities and companies.

**Introduction**

The article that follows is going to define at first the context of the experiment of the action learning carried out in the University of Angers since 1998 and relating to nearly a hundred individual projects.
The targeted public is twofold: administrative staff, technical and since 2005 teachers from University of Angers on one hand, as well as employees of companies and administrations of the region of Angers on the other hand.
The trainees have in common their methodological and technical misunderstanding of requirements needed for the conception of a database. These requirements will be explained later. The challenge, in fact, consists in training people who are not specialists in information systems, or even general data processing.
I will thus explain the educational and organizational choices to adapt the training for the public concerned.
The last part of this article will attempt to show the organization developed to improve the quality of the action learning, the indicators set up, to notice the results obtained.
I will conclude this study with the perspectives of development of this experiment.

**Lifelong learning.**

The Lifelong learning center of the University of Angers (CUFCo) performs a mission of training adults throughout their professional life. The CUFCo proposes modules validated by a certification, as well as degree courses from High school diploma level up to the Master degree preparing to work for a company, named Professional Master.
The aim of this training plan is to allow the acquisition of new skills in order to complete the current course of the trainee, as well as to give the opportunity to start a new career.
The modules of training related to the personal enrichment also concern the voluntary sector and retired persons.
The vocational training of the staffs of the University in computing performed by a private company was attributed to the CUFCo in 1996.
The vocational training presents the following difficulties: the trainees often have different courses, different levels of study and very different work experience. The pedagogy of the trainer will have to adapt to meet the often different expectations of each of them.

Furthermore, if an objective of result is agreed concerning the professional project of every trainee then the trainer has to introduce real individual phases of training alternating with more conventional phases of group training.

**The experience of Consultant.**

The practice of an activity whose financial profit is subjected to the quality of the service makes the implementation of performance indicators essential. It is a question of existence of the company.
We are talking about making profitable the training duration granted by the trainee, and the financial and organizational effort invested by Decisional authority of the trainee.
Introduction of half-time professionals within the teaching staff of the French universities is made possible under status of Associated Teachers ( PAST).

Participation of professionals in the teaching and organizational activities in French university remain marginal. It however contributes to link French universities with the companies likely to recruit the students coming from these same universities.
For the associated professor, coherence and synergy between the university education work and the professional practice are undeniable, particularly if teachers work in a professional institute such as the University Institutes of Technology named IUT or Engineering schools.

**The role of a referent professor.**

University Institutes of Technology (IUT) have existed in France for now 42 years. They make it possible for young people to integrate a University course for a short duration (2 years) after the high school diploma called Baccalaureat, obtained without redoubling at the age of 17 - 18 years old.
The Technological University degree (DUT: 1 year less a Beng), allows approximately 80 % of the students (84% for the Mechanical and Manufacturing Department named GMP Department of Angers-Cholet [1]), as an average in all the specialties proposed by IUT, to continue their studies in engineering schools, in university or in professional License (LP: License level) which is the equivalent of a BEng.
Teaching is based on theoretical courses, tutorial class and a significant percentage of practical classes during which the students apply the knowledge acquired in restricted groups.

Let us take the example of the Mechanical and Manufacturing Department of Cholet, founded in 1996. Students have the possibility of working in autonomy during their 2 (DUT) or 3 (LP) year course on the same machines with numerical control that they will find further in companies.
Teachers of IUT have an evolution in their role during the students courses:
They have to make the students work by themselves during practical work. Then, they assume the role of tutors, during the pedagogical projects that mark out semesters of technological courses, and finally teachers of IUT become referent professors for the 10 to 13 week work placements in companies, which finalize the 2-year course of DUT, and 16 to 20 weeks for the diploma of LP.
Teacher of IUT thus have the possibility of refining the quality of their accompaniment near the student, between support and autonomy.

**Pedagogical Choices:**

Of course spreadsheets-graphics application packages such as Excel or Calc allow to manage lists or data tables, but no databases. (Organization of interconnected tables so as to most accurately model the reality of the information system to be conceived).
It is thus a question of training non data processing specialists with a practice initially reserved for the specialists, the present article wishes to show that teaching the conception of a database will succeed in vocational training only if good tools are given to the trainees:

1. A software like a Relational Database Management System named RDBMS.

2. A standard method of Analysis and Design of .Information Systems named.

3. Documentations with simple and effective contents, likely to be adapted to each trainee.

**Choice of the software packages.**

The software packages Access and Base are implanted in many companies and administrations in France, owing to the fact that they belong multi-field to packs (Office and Open office). Thus the use of the module of database management does not represent any additional cost.
Thanks to this type of software, the company will avoid the pitfalls of the modular solution of the Company Resource Planning named ERP (management of the project too tiresome, software solution not easily adaptable to specificities of the company, and thus too expensive, particularly in terms of training).
Another choice consists in resorting to a Computing Service Company in order to conceive a product by means of a programming language, which will have certainly the advantage to meet the expressed needs.
The major disadvantage will be the subjection of the company to the provider:

- Experience shows that too often the result of the project does not match the expectations of the addressees of the database. In this case, it will be a question of defining the responsibility in financial terms and about deadlines for the updates which will inevitably arise.

- If the design meets the needs stipulated in the specifications, the use of new functionalities and the satisfaction which they cause will reveal needs which were not expressed during the initial analysis.(requests of design of plug-ins to complete the initial project).

**The essential necessity of applying a method of Analysis and Design of Information Systems.**

The fact that module of Database Management System is included in many versions of the Microsoft packs and Open Office ended in an opening of databases to a large public, but also generated numerous unsuccessful experiments.

The other software proposed in the Microsoft and Open Office packs require only some technical knowledge except in the case of users who want or need to use programming. However, creating a database consists in conceiving a data-processing project to satisfy initial specifications. It will require in this particular case the following steps:

- Creating tables of data connected to each other

- Designing tools (fill-in forms, requests called queries, reports) allowing to exploit the database in order to ensure its management (addition, update, suppression of data), to obtain statistics or to lead to structures of exportable data on other software configurations.

It could then mean that Database Management System software packages are only intended for Information software or programming specialists put within the hands of "profane" users.

Another vision of databases makes it possible to consolidate the idea of an inadequacy of the product compared with the level of comprehension of the users (Fig 1):

A professional project of database can come off if it is created by means of Excel, as far as the designer masters the algorithmic and the object-oriented programming language: Visual Basic. Furthermore, the more inconsistent the data is ordered, the heavier the programming is, since the programmer will have to manage with the defects of the phase of analysis.

The second way to obtain a professional project of database consists in using Database Management System software: no programming code is needed, neither in SQL (language of requests of databases) nor in Visual Basic. As there is no miracle, the designer will have to make beforehand an analysis which will need to be perfect.

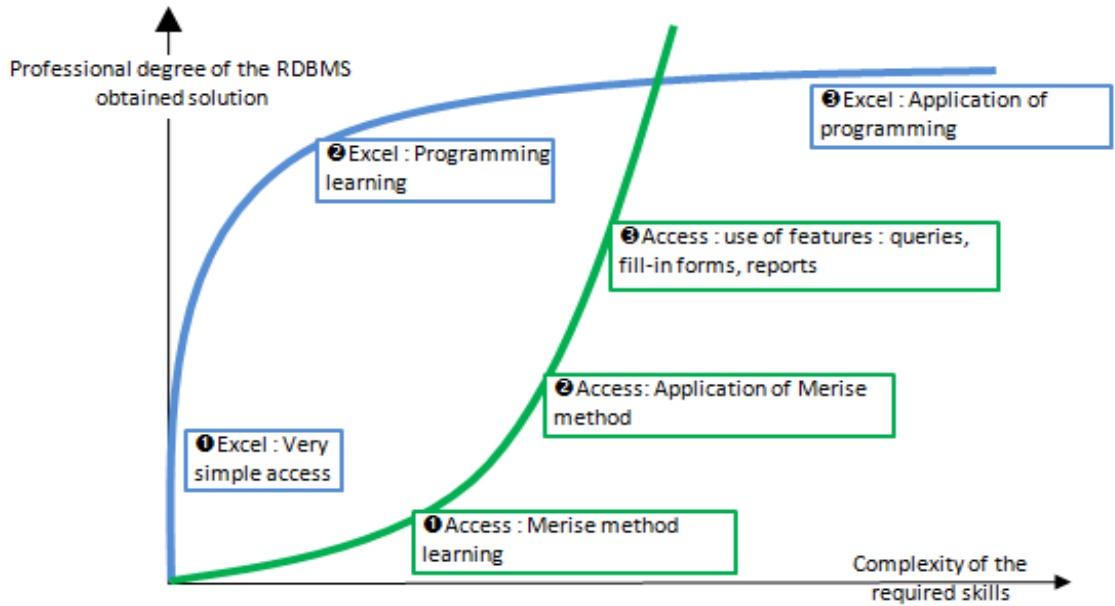

Fig 1 : Diagram to compare two solutions for a Database project : Merise method + Access versus Excel + Programming

**The method of analysis and design : Merise.**

Merise method appeared in France in 1979 as an answer to the request of the Ministry of Industry and was designed by the Computing Technical Center CIT [2] as a method of analysis and conception of information system which would allow to manage large-scale projects within administration and French industries.

The publication of Peter Chen [3] in 1976 allowed to give a name to the bases of the formalism of information systems for next 30 years: the Entity-Association model named EA model.

The Merise method allows to throw a bridge between the needs of the users and data processing specialists, while supplying the specialists with the tool needed for the design of the project, from information to computing.

Its vocation is double :

- Merise is a method for designing information system. It proposes a framework that uses phases of development (Fig 2) which are followed in a logical way, graphic modeling, to make it possible to manage in parallel the analysis and the design of data and processing.

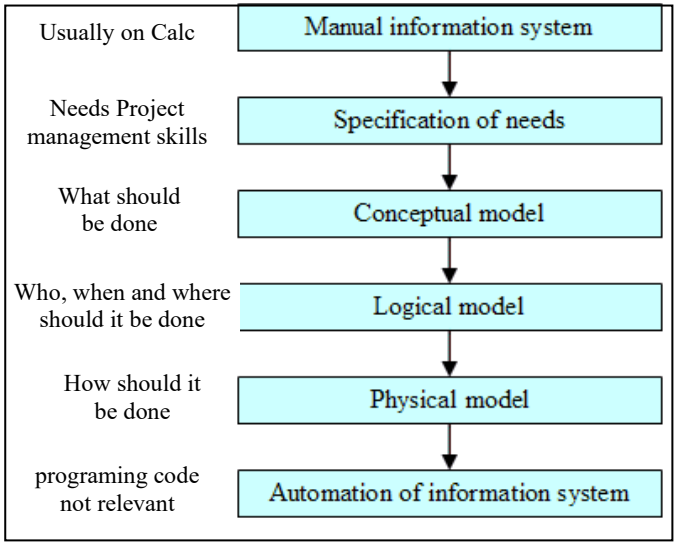

Fig 2 : Merise cycle

- Merise method is an approach for the development of information systems. The recommendations of the method in terms of definition of the working structures on the project, the division of the process of development in steps marked out by phases of communication and validation allow to envisage the whole life cycle and the cycle of decision as a solid base of project management.

The Merise method wanted to be a " methodical trunk " from where the image of the wild cherry (litteral meaning of the term Merise in French) *which can carry beautiful fruits only if a branch of cherry tree is grafted to it*) whose aim was to facilitate new evolutions ensuring its own everlastingness.

Merise is on a phase of decline, gradually replaced by the method Unifed Modeling Language (UML) more adapted to the projects of small and average scale, and definitively directed towards a object modeling. The fields of predilection of the object approach consist in technical and industrial data processing characterized by the management of physical components of the real world.

Merise method applied dogmatically is not adapted for smaller projects developed during the experiment of training described in this article. It proposes however a framework of reflection, an approach by essential steps to make it possible for non-specialists to design themselves the information system they use.
Exploited in the forms of simplified but not denatured joined steps so as to create a coherence, Merise method appears to be particularly effective, while keeping its strong pedagogical potential.

Unified Modeling Language Method, definitively centered on the treatment aspects, fully satisfies data processing specialists who wish to implement their project with object-oriented programming language.
If the objective of the training is to supply all the necessary means in order to allow non- specialists to acquire the project ownership of their own project, use of programming codes will obviously be excluded from the pedagogy selected for this experiment.(Fig 3)

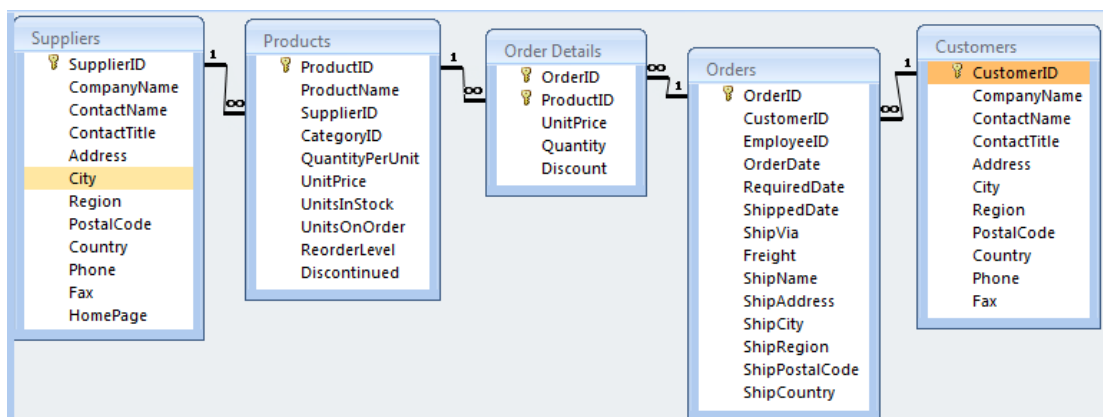

Fig 3 : An example of database structure

**Choice and exploitation of the pedagogical contents**

The educational contents may allow the teacher to:
- Adapt to the heterogeneousness of the group
- Facilitate the practice on cases the closest to the professional reality.

The solution exploited during this experiment consisted in creating teaching documentations of self training.
Supports concerning Merise method treat a professional case to illustrate the process, gradually leading to the modeling of the structure of the database. The documentation is marked out by a dozen practical cases

taken from the environment industrial as well as from services and administrations

The corrections are made in common, for a number of the cases suggested, and this to gather regularly all the trainees so that they have the possibility of thinking in team about a common project.

The additional case studies form an available bank for the trainees who may progress faster.

Supports concerning Database Management System software are entirely used in a autonomous way. A first documentation allows trainees to obtain a brief vision of all the compartments of the creation and the exploitation of database applications that trainees are going to work on.

Then, the case study used for the Merise method is used there too as support for the part discovered of each module of course (creation of the tables, connection of the database, queries, fill-in forms, reports.). Each of these documentations is marked out by exercises of progressive difficulties: at first, an exercise similar to the work detailed in the course, then a more opened exercise, requiring from the trainee to succeed in extracting himself from the documentation.

At this stage of the training, each trainee follows his or her own rhythm of progression. The teacher practices an individual pedagogy. He will not forsake therefore the phases of regrouping, which he can set up regularly to exploit a common time of reflection on a more theoretical part (mostly related to Merise method), or concerning a technical point of which trainees would not have perceived the importance.

It is true that the supports of self-training can create a harmful side effect: the trainee follows the detailed indications, and forgets to stand back from the learning that he is acquiring. Regroupings for common reflections are a first answer. A regular experimentation of newly acquired knowledge on a personal professional project for each trainee ensures the teacher of the good understanding of each one.

Thus self-training documentations if well used, make it possible for the teacher to be present for each trainee when they need some help, and for everybody in order to preserve the coherence of the group.

**Planning of an Action Learning**

The training evolved every year with the cooperation of the CUFCo and the person in charge of lifelong learning for the employees and teachers of the University of Angers throughout this experiment. The objective was to supply a learning to be the most effective possible, in a very limited duration. The action learning is an educational choice, which alternates conventional training with individual phases of improvement in professional project environment: the database project each trainee has planned to work on.

Every stage of the modeling of the database with Merise method, or of the physical conception of the database on software, is an opportunity to test knowledge recently acquired on the personal professional project.
The trainer takes then the consulting role in order to ensure a satisfactory progression.

Training went from one to three modules (Fig 4):

- Common training for 6 weeks allowing every trainee to leave with a modeling carried out by their own means and validated by the consulting trainer (or a core of coherent project when the personal professional project is too vast for a first experience, and within the time given)..

- Project-monitoring phase concerns the trainees outside the the university belonging to an organization likely to finance a complete training plan. The project monitoring planned is individual, based on half days of consulting in the trainee office, for specific and very targeted training if it is necessary, with remote supply ensured during all the phase of the project-monitoring.

- The last module, articulated in the same manner as the first training module, concerns the trainees who want to make their database evolve, to reach complementary knowledge. Some projects too vast to be modeled and designed during the first module, can be the object of extensions, which are added to the basic coherent core of initial project. This last phase gives the trainee the opportunity to make sure that they master the process of design, by summing all the steps of the process

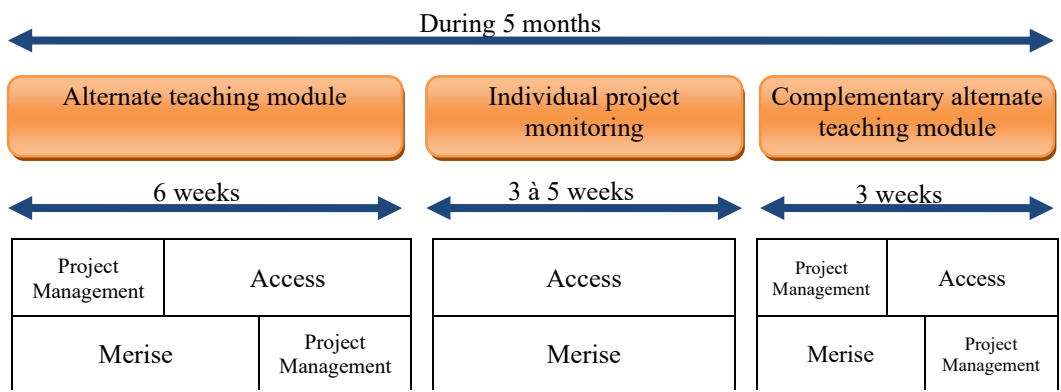

Fig 4 : Action learning planning

**Evaluation of the experiment and prospective**

10 years of experimentation made allowed to make evolve the training until its actual organization.

Some difficulties appeared:
Needs assessment for the most part of the projects was not clearly established. This situation created disappointments concerning the result. The module of project management allowed resolving this part of dissatisfaction.
The module of project management gave to the trainees the possibility to establish the specifications of their project. Specifications give the possibility to contractualize the objectives of the training between the trainer consultant and the trainee.

Trainees are ensured to finish the 1rst phase named common training with the structure of their database:
- Designed thanks to Merise method,
- Certified valid by the trainer in coherence with specifications,
- Implemented (tables connected) on the software.

Some restrictions concern vast projects, but this problem is well managed with use of the core of coherent project.
Another point gives troubles: some trainees would like to finalize the application implementation with every tools (queries, fill-in forms and reports) they intended to create. Project monitoring module and improvement module could give to the trainee the opportunity to reach new ambitious objectives.

The evaluation of the training gives a result of about 85% of satisfaction concerning the obligation of means, and 70% concerning the obligation of result.

Prospectives for this training :
- Try to insist even more on the project management aspect, contractualization of the objectives, and formalization of specifications.
- Adapt the pedagogical organization in remote access related to a teaching process of e-learning.

**Conclusions**

The training described in this article proposes services which nor training center neither Computing Service Company are able to provide :

No training center in the area of Angers would take the risk to give to the trainees the opportunity to work on their own professional project, because the heterogeneity of the projects to treat added to the heterogeneity related to each trainee, is too difficult to be managed during a training period.

However, the aim of the action learning associated with pedagogical adapted tools such as the design and the use of documentation of self-training make it possible to render the double service: training and especially consulting.

No Computing Service Company in the area of Angers proposes to its customers to transfer totality of its competences. If it is however the case, the design is carried out with programming code, either by preoccupations with effectiveness for the consultant, or by the will to assure an exclusivity for futher developpements on the project.

No training center in the area of Angers announces obligation of result to its customers

The training described in this article clearly shows the will of a complete transfer of competences. The results of trainee satisfaction survey shows that the consultant trainer increases trainee satisfaction, and by the way fidelize the customer for futher project monitoring, new evolutions of the project, expertise of developments assumed by the trainee. The decision maker is often interested by the double objective to obtain a concrete result while allowing to his employees the opportunity to acquire new skills.
A project conceived by an employee belonging to the consumer organization is definitely less expensive than an application created by a Computing Service Compagny.